\documentclass[twocolumn]{aastex63}
\shorttitle{SOFIA/HAWC+ view of an ELIRG}
\shortauthors{Toba et al.}

\begin{document}
\title{SOFIA/HAWC+ View of an Extremely Luminous Infrared Galaxy, WISE1013+6112}

\correspondingauthor{Yoshiki Toba}
\email{toba@kusastro.kyoto-u.ac.jp}

\author[0000-0002-3531-7863]{Yoshiki Toba}
\affiliation{Department of Astronomy, Kyoto University, Kitashirakawa-Oiwake-cho, Sakyo-ku, Kyoto 606-8502, Japan}
\affiliation{Academia Sinica Institute of Astronomy and Astrophysics, 11F of Astronomy-Mathematics Building, AS/NTU, No.1, Section 4, Roosevelt Road, Taipei 10617, Taiwan}
\affiliation{Research Center for Space and Cosmic Evolution, Ehime University, 2-5 Bunkyo-cho, Matsuyama, Ehime 790-8577, Japan}

\author[0000-0003-2588-1265]{Wei-Hao Wang}
\affiliation{Academia Sinica Institute of Astronomy and Astrophysics, 11F of Astronomy-Mathematics Building, AS/NTU, No.1, Section 4, Roosevelt Road, Taipei 10617, Taiwan}

\author[0000-0002-7402-5441]{Tohru Nagao}
\affiliation{Research Center for Space and Cosmic Evolution, Ehime University, 2-5 Bunkyo-cho, Matsuyama, Ehime 790-8577, Japan}

\author[0000-0001-7821-6715]{Yoshihiro Ueda}
\affiliation{Department of Astronomy, Kyoto University, Kitashirakawa-Oiwake-cho, Sakyo-ku, Kyoto 606-8502, Japan}

\author{Junko Ueda}
\affiliation{National Astronomical Observatory of Japan, 2-21-1 Osawa, Mitaka, Tokyo, 181-8588, Japan}

\author{Chen-Fatt Lim}
\affiliation{Graduate Institute of Astrophysics, National Taiwan University, PO Box 23-141, Taipei 10617, Taiwan}
\affiliation{Academia Sinica Institute of Astronomy and Astrophysics, 11F of Astronomy-Mathematics Building, AS/NTU, No.1, Section 4, Roosevelt Road, Taipei 10617, Taiwan}

\author[0000-0002-6720-8047]{Yu-Yen Chang}
\affiliation{Academia Sinica Institute of Astronomy and Astrophysics, 11F of Astronomy-Mathematics Building, AS/NTU, No.1, Section 4, Roosevelt Road, Taipei 10617, Taiwan}

\author[0000-0002-2501-9328]{Toshiki Saito}
\affiliation{Max Planck Institute for Astronomy, K\"onigstuhl 17, D-69117 Heidelberg, Germany} 

\author[0000-0002-8049-7525]{Ryohei Kawabe}
\affiliation{National Astronomical Observatory of Japan, 2-21-1 Osawa, Mitaka, Tokyo, 181-8588, Japan}
\affiliation{Department of Astronomy, The University of Tokyo, 7-3-1 Hongo, Bunkyo-ku, Tokyo 113-0033, Japan}
\affiliation{Department of Astronomical Science, Graduate University for Advanced Studies (SOKENDAI), 2-21-1 Osawa, Mitaka, Tokyo 181-8588, Japan}



\begin{abstract}
We present far-infrared (FIR) properties of an extremely luminous infrared galaxy (ELIRG) at $z_{\rm spec}$ = 3.703, WISE J101326.25+611220.1 (WISE1013+6112).
This ELIRG is selected as an IR-bright dust-obscured galaxy (DOG) based on the photometry from the Sloan digital sky survey (SDSS) and {\it wide-field infrared survey explorer (WISE)}. 
In order to derive its accurate IR luminosity, we perform follow-up observations at 89 and 154 $\micron$ using the high-resolution airborne wideband camera-plus (HAWC+) on board the 2.7-m stratospheric observatory for infrared astronomy (SOFIA) telescope.
We conduct spectral energy distribution (SED) fitting with {\tt CIGALE} using 15 photometric data (0.4--1300 $\micron$).
We successfully pin down FIR SED of WISE1013+6112 and its IR luminosity is estimated to be $L_{\rm IR}$ = (1.62 $\pm$ 0.08) $\times 10^{14}$$L_{\sun}$, making it one of the most luminous IR galaxies in the universe.
We determine the dust temperature of WISE1013+6112 is $T_{\rm dust}$ = 89 $\pm$ 3 K, which is significantly higher than that of other populations such as SMGs and FIR-selected galaxies at similar IR luminosities.
The resultant dust mass is $M_{\rm dust} = (2.2 \pm 0.1) \times 10^{8}$ $M_{\sun}$.
This indicates that WISE1013+6112 has a significant active galactic nucleus (AGN) and star-forming activity behind a large amount of dust.
\end{abstract}

\keywords{galaxies: active --- infrared: galaxies --- (galaxies:) quasars: supermassive black holes --- (galaxies:) quasars: individual (WISE J101326.25+611220.1}


\section{Introduction}

Galaxies whose infrared (IR) luminosity exceeds 10$^{13}$$L_{\sun}$ and 10$^{14}$$L_{\sun}$ have been termed as hyper-luminous IR galaxies \citep[HyLIRGs:][]{Rowan-Robinson} and extremely-luminous IR galaxies \citep[ELIRGs:][]{Tsai}, respectively.
Their IR luminosity ($L_{{\rm IR}}$) is expected to be produced by star formation (SF), active galactic nucleus (AGN) activity, or both.
In the context of major merger scenario, their extreme IR luminosity could indicate that it corresponds to the peak of AGN and/or SF activity behind a large amount of gas and dust \citep{Narayanan,Ricci,Blecha}.
Therefore, it is important to search for IR luminous galaxies such as HyLIRGs and ELIRGs for understanding the galaxy formation and evolution and connection to their super massive black holes (SMBHs) \citep[see e.g.,][]{Hopkins}.
However, their volume densities are extremely low \citep{RW,Gruppioni}, and thus wide and deep surveys are required to detect these spatially rare populations.

One successful technique to search for HyLIRGs and ELIRGs is based on mid-IR (MIR) colors taken with {\it Wide-field Infrared Survey Explorer} \citep[{\it WISE}:][]{Wright}.
Most of objects that are faint or undetected by {\it WISE} at 3.4 $\micron$ (W1) and 4.6 $\micron$ (W2) but are well detected at 12 $\micron$ (W3) or 22 $\micron$ (W4) are classified as HyLIRGs/ELIRGs.
They are termed hot dust-obscured galaxies (DOGs\footnote{The original definition of DOGs was flux density at 24 $\micron > 0.3$ mJy and $R$ --[24] $>$ 14, where R and [24] represent Vega magnitudes in the $R$-band and 24 $\micron$, respectively \citep[see][for more detail]{Dey}.}) or ``W1W2 dropouts'' \citep{Eisenhardt,Wu}.
Indeed, a hot DOG with $L_{{\rm IR}} = 2.2 \times 10^{14}\,L_{\sun}$ was reported as the most luminous galaxy in the universe \citep{Tsai}.
However, \cite{Fan_18} recently reported that this ELIRG is contaminated by a foreground galaxy, resulting in an
over estimation of its total IR luminosity by a factor of about two \citep[see also][]{Tsai_18}.

\cite{Toba_16} also performed an extensive search for HyLIRGs and ELIRGs by using the Sloan digital sky survey \citep[SDSS:][]{York} and {\it WISE}.
By combining the SDSS Data Release 12 \cite[DR12:][]{Alam} spectroscopic catalog and ALLWISE catalog \citep{Cutri}, they selected optically-faint but IR bright objects with $i - [22] > 7.0$ and flux density at 22 $\micron$$>$ 3.8 mJy in 14,555 deg$^2$, where $i$ and [22] are $i$-band and 22 $\mu$m AB magnitudes, respectively, yielding 67 objects with spectroscopic redshift.
These objects are known as IR-bright DOGs \citep{Toba_15,Toba_17,Noboriguchi}.
\cite{Toba_16} then estimated their tentative IR luminosities based on the spectral energy distribution (SED) fitting with a SED fitting code. SED analysis using Bayesian statistics \citep[{\sf SEABASs};][]{Rovilos} \citep[see also][]{Toba_17c}, where they used only SDSS and {\it WISE} data (see Figure 6 in \citealt{Toba_16}).
Consequently, an IR-bright DOG, WISE J101326.25+611220.1 (hereafter WISE1013+6112) at spectroscopic redshift ($z_{\rm spec}$) = 3.70, was left as an ELIRG candidate.

\cite{Toba_18} then executed follow-up observations of WISE1013+6112 with the submillimetre common user bolometer array 2 \citep[SCUBA-2:][]{Holland} on the James Clerk Maxwell telescope (JCMT) (S17AP002, PI: Y.Toba), and the Submillimeter Array \citep[SMA:][]{Ho} (2016BA003, PI: Y.Toba).
They performed the SED fitting by adding data points at 450 and 850 $\micron$ (SCUBA-2/JCMT) and 870 and 1300 $\micron$ (SMA).
The derived IR luminosity was $L_{\rm IR} = 2.2^{+1.5}_{-1.0} \times 10^{14}$$L_{\sun}$, making it an ELIRG.
However, as we did not have deep rest-frame MIR and far-IR (FIR) photometry responsible for FIR SED, the derived IR luminosity remains a large uncertainty.
In order to constrain IR luminosity of this ELIRG more accurately and to investigate the SF activity and dust property of its host galaxy, we require deep FIR data.

In this paper, we present follow-up observations of an extremely luminous DOG, WISE1013+6112, at 89 and 154 $\micron$ using a high-resolution airborne wideband camera-plus \citep[HAWC+:][]{Harper} on the 2.7-m stratospheric observatory for infrared astronomy (SOFIA) telescope \citep{Temi}.
These observations with HAWC+/SOFIA enable us to pin down the FIR-SED of WISE1013+6112.
Throughout this paper, the adopted cosmology is a flat universe with $H_0$ = 70 km s$^{-1}$ Mpc$^{-1}$, $\Omega_M$ = 0.3, and $\Omega_{\Lambda}$ = 0.7, which are same as those adopted in \cite{Toba_18}.
Solar luminosity is defined as $L_{\sun} = 3.828 \times 10^{33}$ erg s$^{-1}$ \citep{Mamajek}.

\section{Data and analysis}

\subsection{Follow-up observations with SOFIA}

\begin{figure}
    \centering
    \includegraphics[width=0.45\textwidth]{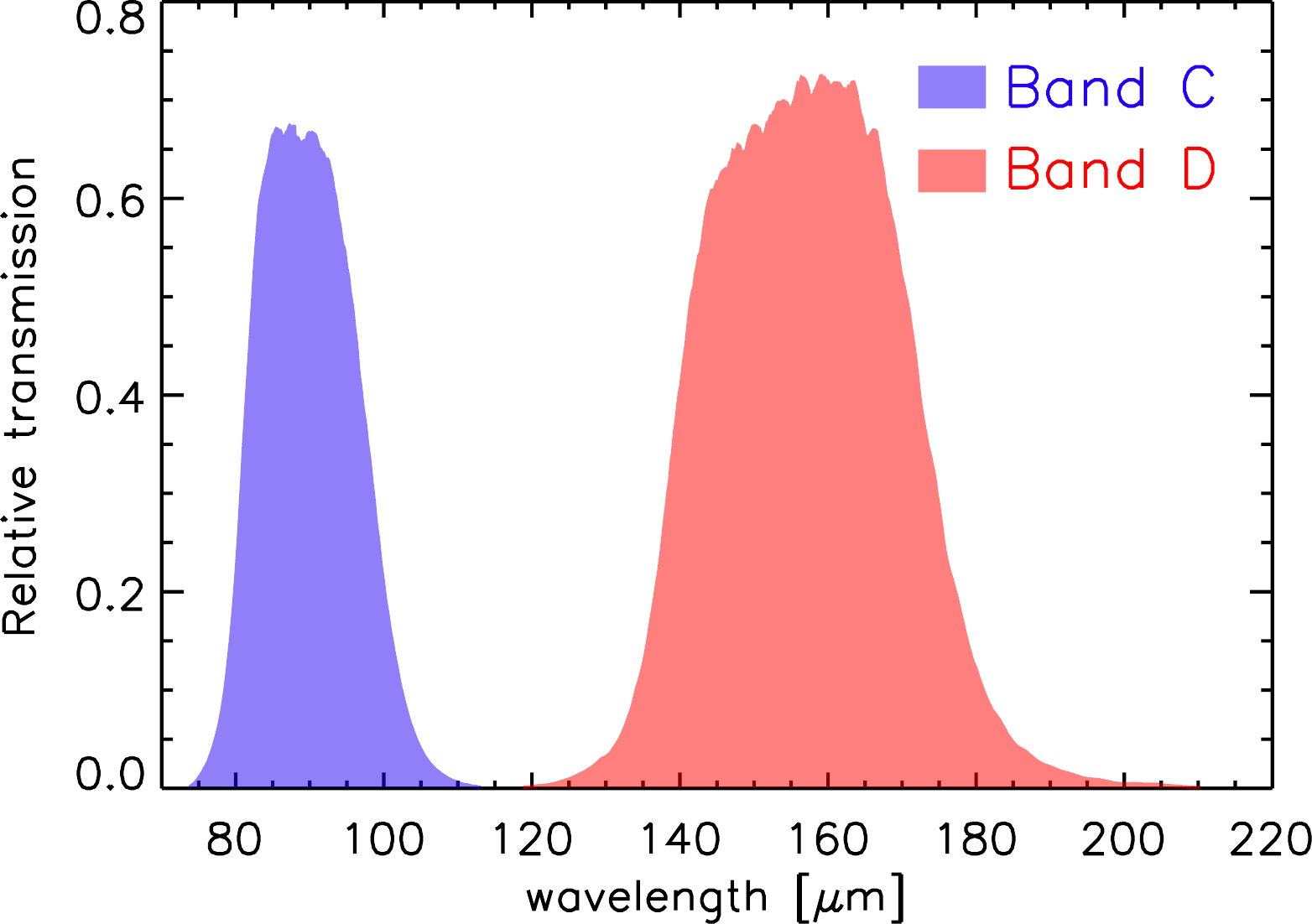}
\caption{HAWC+ filter transmission profiles for band C (blue) and D (red).}
\label{filter}
\end{figure}

Flux densities at 89 and 154 $\micron$ were obtained using HAWC+/SOFIA in Cycle 6 (PI: Y.Toba).
The filter transmission profiles\footnote{\url{https://www.sofia.usra.edu/science/proposing-and-observing/observers-handbook-cycle-8/7-hawc/71-specifications\#FiltersHAWC}} for these bands are shown in Figure \ref{filter}.
Data have been obtained for the observation (PlanID: 06\_0029) during HAWC+ mission 2019-02-13\_HA\_F546.
We performed the total intensity mapping with HAWC+ bands C (89 $\micron$) and D (154 $\micron$) providing angular resolutions of 7$\arcsec$.8 and 13$\arcsec$.6 in full width at half maximum (FWHM), respectively.
The total on-source integration times were approximately 100 minutes at both 89 and 154 $\micron$. 
Data were reduced using the HAWCDPR PIPELINE v1.3.0 \citep{Harper}.
As the source is faint, this data is processed using the faint option.

\begin{figure*}
    \centering
    \includegraphics{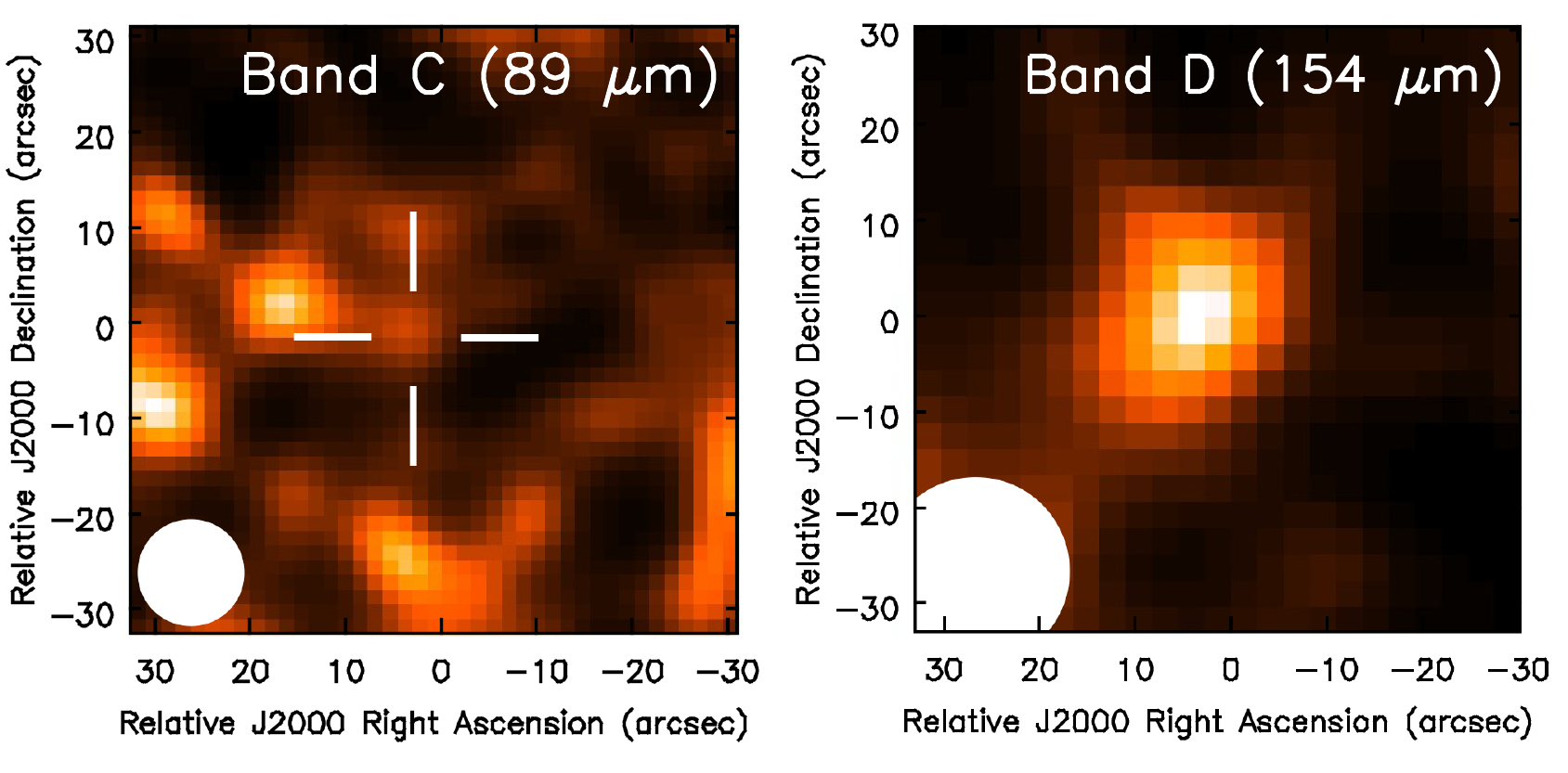}
\caption{Total flux images at 89 $\micron$ (left) and 154 $\micron$ (right). Relative coordinate in units of arcsec with respect to the SDSS position of WISE1013+6112 is employed. The white filled circles are beam sizes for each band. The object considered here in the left panel is located at the center of the white cross.}
\label{image}
\end{figure*}

Figure \ref{image} shows the FIR image of WISE1013+6112 taken by HAWC+.
WISE1013+6112 was marginally detected at 89 $\micron$ with signal-to-noise ratio (S/N) = 4.1 while this object was clearly detected at 154 $\micron$ with S/N = 9.8 that were evaluated using 2D Gaussian fitting (see below).
The flux measurements were performed in the same manner as performed for SCUBA-2 and SMA data in \cite{Toba_18}.
We employed the common astronomy software applications package \citep[CASA ver. 5.5.0;][]{McMullin}.
We performed a 2D Gaussian fit for each image and estimated the total fluxes within 10$\arcsec \times 10\arcsec$ and 20$\arcsec \times 20\arcsec$ aperture, respectively.
The photometry of WISE1013+6112 including SOFIA FIR flux densities measured in this work are summarized in Table \ref{Table}.

\begin{table}
\renewcommand{\thetable}{\arabic{table}}
\centering
\caption{Observed Properties of WISE1013+6112}
\label{Table}
\begin{tabular}{lr}
\tablewidth{0pt}
\hline
\hline
WISE J101326.25+611220.1		&								\\
\hline
R.A. (SDSS) [J2000.0] 			&	10:13:26.24					\\
Decl. (SDSS) [J2000.0]			&	+61:12:19.76  				\\
Redshift (SDSS)					&	3.703 $\pm$ 0.001			\\
SDSS$u$-band [$\mu$Jy]	&	$<$ 1.26\tablenotemark{a}					\\
SDSS$g$-band [$\mu$Jy]	&	3.47 $\pm$ 0.47				\\
SDSS$r$-band [$\mu$Jy]	&	13.70 $\pm$ 0.67			\\
SDSS$i$-band [$\mu$Jy]	&	13.58 $\pm$ 0.95			\\
SDSS$z$-band [$\mu$Jy]	&	21.09 $\pm$ 4.03			\\
{\it WISE} 3.4 $\micron$ [mJy]	&	0.05 $\pm$ 0.01				\\
{\it WISE} 4.6 $\micron$ [mJy]	&	0.13 $\pm$ 0.01				\\
{\it WISE} 12  $\micron$ [mJy]	&	3.30 $\pm$ 0.16				\\
{\it WISE} 22  $\micron$ [mJy]	&	10.70 $\pm$ 0.98			\\
HAWC+/SOFIA 89 $\micron$ [mJy]	&	22.5 $\pm$ 5.5			\\
HAWC+/SOFIA 154 $\micron$ [mJy]	&	63.4 $\pm$ 6.5			\\
SCUBA-2/JCMT 450 $\micron$ [mJy]		&	46.00 $\pm$ 8.05\tablenotemark{b}			\\
SCUBA-2/JCMT 850 $\micron$ [mJy]		&	13.35 $\pm$ 0.67\tablenotemark{b}			\\
SMA 870 $\micron$ [mJy]			&	13.60 $\pm$ 2.72\tablenotemark{b}			\\
SMA 1.3 mm [mJy]				&	6.49 $\pm$ 1.30\tablenotemark{b}				\\
$L_{\rm IR}$ [$L_{\sun}$]			&	$(1.62 \pm 0.08) \times 10^{14}$	\\
$L^{\rm AGN}_{\rm IR}$ [$L_{\sun}$]	&	$(1.13 \pm 0.06) \times 10^{14}$	\\
$L^{\rm SF}_{\rm IR}$ [$L_{\sun}$]	&	$(0.49 \pm 0.10) \times 10^{14}$	\\
$M_{\rm *}$ [$M_{\sun}$]								&	$(2.03 \pm 0.36) \times 10^{11}$	\\
SFR [$M_{\sun}$ yr$^{-1}$]								&	$(2.81 \pm 0.36) \times 10^{3}$		\\
$T_{\rm dust}$ [K]										&	$(8.9 \pm 0.3) \times 10$			\\
$M_{\rm dust}$ [$M_{\sun}$]								&	$(2.2 \pm 0.1) \times 10^{8}$	\\
\hline
\multicolumn{2}{l}{(a) 3$\sigma$ upper limit.}\\
\multicolumn{2}{l}{(b) see \cite{Toba_18} in details.}
\end{tabular}
\end{table}

\subsection{SED fitting with {\tt CIGALE}}
\label{CIGALE}

We employed {\tt CIGALE}\footnote{\url{https://cigale.lam.fr/2018/11/07/version-2018-0/}} \citep[code investigating galaxy emission:][]{Burgarella,Noll,Boquien} to conduct a detailed SED modeling in a self-consistent framework by considering the energy balance between the ultraviolet/optical and IR. 
In this code, users can handle various parameters, such as star formation history (SFH), single stellar population (SSP), attenuation law, AGN emission, dust emission, and radio synchrotron emission \cite[see e.g.,][]{Toba_19a,Toba_19b,Toba_19c}.

SFH is assumed as two exponential decreasing star formation rate (SFR)with different e-folding times \citep{Ciesla_15,Ciesla_16}, where we parameterized e-folding time of the main stellar population ($\tau_{\rm main}$) and the late starburst population ($\tau_{\rm burst}$), mass fraction of the late burst population ($f_{\rm burst}$), and age of the main stellar population in the galaxy \citep[see Section 3.1.2 in][in details]{Boquien}. 
We used the stellar templates provided from \cite{Bruzual} assuming the \cite{Chabrier} initial mass function (IMF), and the standard default nebular emission model included in {\tt CIGALE} \citep[see ][]{Inoue}.
Dust attenuation is modeled by using the \cite{Calzetti} starburst attenuation curve with small Magellanic cloud (SMC) extinction curve \citep{Pei}, where the color excess of the emission lines $E(B-V)_{\rm lines}$ is parameterized.  
The color excess of stars, $E(B-V)_{*}$ can be converted from $E(B-V)_{\rm lines}$ by assuming a simple reduction factor ($f_{\rm EBV}$ = $\frac{E(B-V)_{*}}{E(B-V)_{\rm lines}}$) = 0.44 \citep{Calzetti97}.
For AGN emission, we utilized models provided by \cite{Fritz}.
In order to avoid a degeneracy of AGN templates in the same manner as in \cite{Ciesla_15} and \cite{Toba_19b}, we fixed certain parameters that determine the density distribution of the dust within the torus, i,e., ratio of the maximum to minimum radii of the dust torus ($R_{\rm max}/R_{\rm min}$), optical depth at 9.7 $\micron$ ($\tau_{\rm 9.7}$), density profile along the radial and the polar distance coordinates parameterized by $\beta$ and $\gamma$ \citep[see equation 3 in][]{Fritz}, and opening angle ($\theta$).
Hence, we parameterized the $\psi$ parameter (an angle between equatorial axis and line of sight) that corresponds to a viewing angle of the torus.
We further parameterized AGN fraction ($f_{\rm AGN}$) that is the contribution of IR luminosity from the AGN to the total IR luminosity \citep{Ciesla_15}.

As one of the purposes of this work is to derive the dust temperature ($T_{\rm dust}$), we employed the analytic model provided by \cite{Casey} for dust emission.
This model consists of two components: one is a single temperature modified black body (MBB) and the other is power-law emission in the MIR.
As the MIR power-low component is expected to be dominated by the AGN torus emission that was already taken into account in \cite{Fritz} model, we focus only on the MBB component.
MBB is formulated as $1-e^{\tau (\nu)}$ $\nu^{\beta}\,B_{\rm \nu} (T_{\rm dust})$, where $\nu$ is the frequency, $\beta$ is the emissivity index of the dust, and $B_{\rm \nu} (T_{\rm dust})$ is the Planck function.
$\tau \equiv (\nu/\nu_0)^{\beta}$ is the optical depth, where $\nu_0$ is the frequency where optical depth equals unity \citep{Draine_06}.
In this work, we fixed $\nu_0$ = 1.5 THz ($\lambda_0$ = 200 $\micron$) \citep[e.g.,][]{Conley} and $\beta$ = 1.6 \citep[e.g.,][]{Fan_16}, and parameterized only $T_{\rm dust}$.
We confirmed that the choice of $\nu_0$ and $\beta$ does not significantly affect the following results as long as adopting $\nu_0 < 1.5$ THz and $\beta$ = 1--2 \citep[see also][]{Kovacs}.
Note that \cite{Casey} model does not include polycyclic aromatic hydrocarbon (PAH) emission that could dominate MIR emission (particularly for SF galaxies), in exchange for parameterizing dust temperature.
However, we confirmed that the resultant IR luminosity is consistent with what we reported in this work even when we used other dust models with PAH emission such as the one by \cite{Dale}.
The detailed parameter ranges adopted in the SED fitting is tabulated in Table \ref{Param}.

\begin{table}
\renewcommand{\thetable}{\arabic{table}}
\centering
\caption{Parameter Ranges used in the SED Fitting with {\tt CIGALE}} 
\label{Param}
\begin{tabular}{lc}
\tablewidth{0pt}
\hline
\hline
Parameter & Value\\
\hline
\multicolumn2c{Double exp. SFH}\\
\hline
$\tau_{\rm main}$ [Myr] & 50, 100, 500, 1000, 3000		\\
$\tau_{\rm burst}$ [Myr] & 3, 5, 8, 10 \\
$f_{\rm burst}$& 0.5, 0.7, 0.8, 0.9, 0.95 \\
age [Myr] &  500, 1000, 3000, 5000 \\
\hline
\multicolumn2c{SSP \citep{Bruzual}}\\
\hline
IMF				&	\cite{Chabrier} \\
Metallicity		&	0.02 \\
\hline
\multicolumn2c{Dust attenuation \citep{Calzetti}}\\
\hline
$E (B-V)_{\rm lines}$	&  0.2, 0.4, 0.6, 0.8, 1.0		\\
$f_{\rm EBV}$			& 0.44							\\
Extinction curve		&SMC \citep{Pei}				\\
\hline
\multicolumn2c{AGN emission \citep{Fritz}}\\
\hline
$R_{\rm max}/R_{\rm min}$& 150  \\
$\tau_{\rm 9.7}$& 0.6 \\
$\beta$& 0.00\\
$\gamma$& 0.0 \\
$\theta$& 60\\
$\psi$& 0.001, 60.100, 89.990 \\
$f_{\rm AGN}$& 0.1, 0.2, 0.3, 0.4, 0.5, 0.6, 0.7, 0.8, 0.9 \\
\hline
\multicolumn2c{Dust emission \citep{Casey}}\\
\hline
$T_{\rm dust}$ [K] &	10, 20, 30, 40, 50, 60, 70, 80, 90, 100\\
Emissivity $\beta$	&	1.6\\
\hline
\end{tabular}
\end{table}

\section{Results and Discussions}
\subsection{IR luminosity}
\label{LIR}

\begin{figure*}
    \centering
    \includegraphics{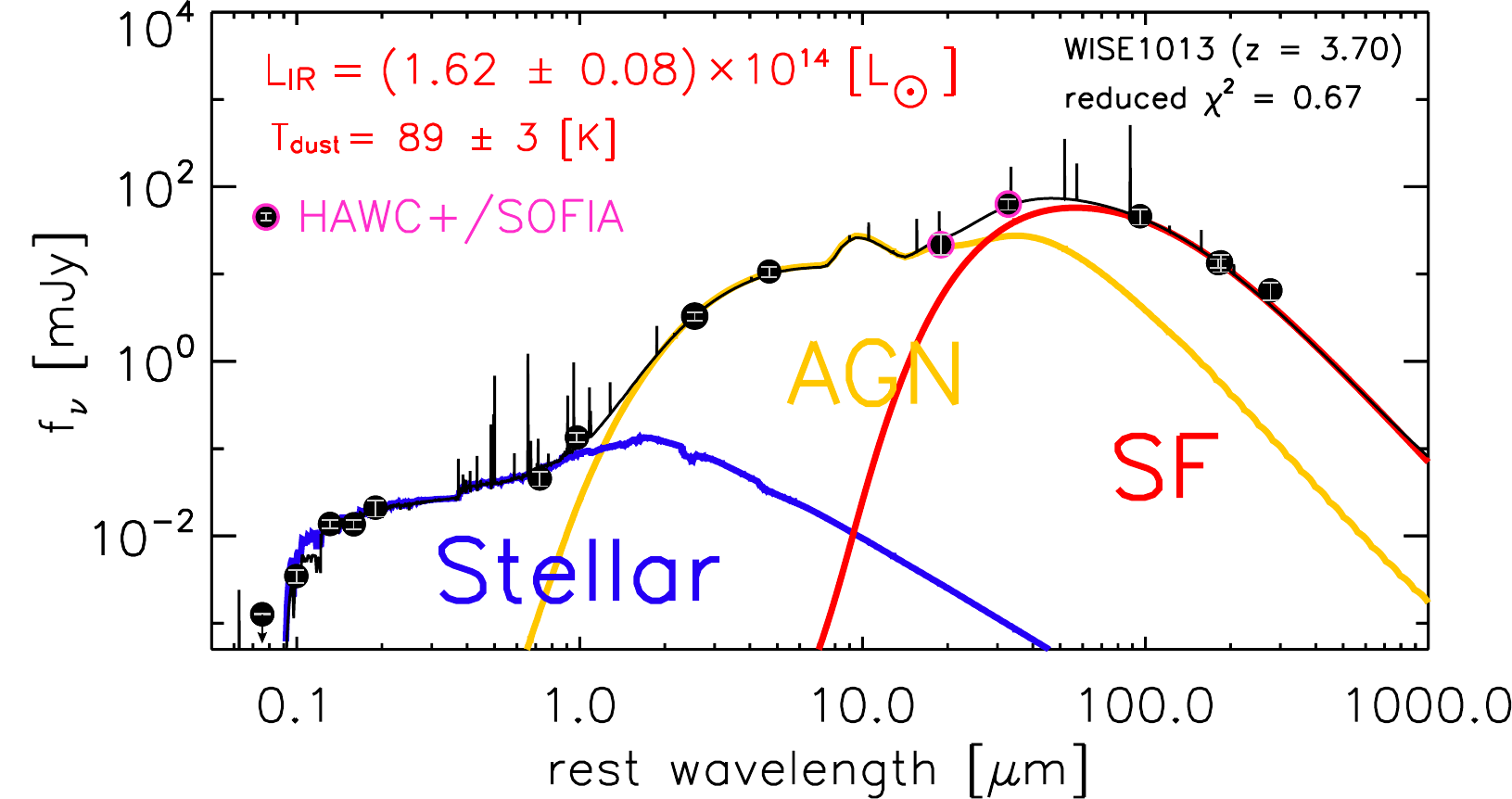}
\caption{SED of WISE1013+6112. The black points are photometric data, two of which are new data obtained in this work (black-magenta circles).
The contribution from the stellar, AGN, and SF components to the total SED are shown as blue, yellow, and red lines, respectively. The black solid line represents the resultant best-fit SED.}
\label{SED}
\end{figure*}

Figure \ref{SED} shows the SED of WISE1013+6112 in the rest frame at $z_{\rm spec}$ = 3.70.
The observed data points of WISE1013+6112 are well-fitted by the combination of stellar, AGN, and SF components with an adequately reduced $\chi^{2}$ (= 0.67).
The resultant IR luminosity is $L_{\rm IR} = (1.62 \pm 0.08) \times 10^{14}$ $L_{\sun}$ that is consistent with that reported in \cite{Toba_18} ($L_{\rm IR} = 2.2^{+1.5}_{-1.0} \times 10^{14}$ $L_{\sun}$) within the errors.
Nevertheless, we have successfully estimated IR luminosity more accurately owing to additional FIR data taken by HAWC+/SOFIA; the relative error of $L_{\rm IR}$ was reduced approximately to 5\%.
We confirmed that WISE1013+6112 is still one of the most IR luminous galaxies in the universe.

The AGN fraction defined as $L_{\rm IR}$\,(AGN)/$L_{\rm IR}$ is 0.7, which is smaller than what was reported in \cite{Toba_18} ($f_{\rm AGN} = 0.9^{+0.06}_{-0.20}$).
This is because \cite{Toba_18} overestimated AGN luminosity owing to the lack of SOFIA data that covered the peak of the FIR luminosity.
We found that WISE1013+6112 is still AGN-dominated but SF luminosity moderately contributes to the total IR luminosity (see Section \ref{host}).

\subsection{Host properties}
\label{host}

We discuss host properties, stellar mass ($M_{*}$) and star formation rate (SFR), of WISE1013+6112 in this subsection.
The resultant $M_{*}$ and SFR outputs by {\tt CIGALE} are $M_{*} = (2.03 \pm 0.36) \times 10^{11}$ $M_{\sun}$ and SFR = $(2.81 \pm 0.36) \times 10^{3}$ $M_{\sun}$ yr$^{-1}$, respectively.
The stellar mass is in good agreement with that reported in \cite{Toba_18}.
On the other hand, SFR is 2.2 times larger than that reported in \cite{Toba_18}.
This is reasonable because \cite{Toba_18} underestimated the SF luminosity as discussed in Section \ref{LIR}, and SFR derived here is more reliable with small uncertainty.
We confirmed that WISE1013+6112 shows a significant offset with respect to the main-sequence (MS) galaxies at $3 < z < 4$ \citep{Tomczak}.
Given the same stellar mass, SFR of WISE1013+6112 is roughly an order of magnitude higher than that of SF galaxies at similar redshifts, suggesting that WISE1013+6112 still has very active star formation.
The resultant SFH, i.e., ($\tau_{\rm main}$, $\tau_{\rm burst}$, $f_{\rm burst}$, age) = (100 Myr, 10 Myr, 0.95, 1000 Myr) suggests that WISE1013+6112 might have an instantaneous starburst that lasts a few hundred Myr.

\subsection{Dust temperature}
\label{dust}

We then discuss the dust temperature ($T_{\rm dust}$) heated by SF activity in WISE1013+6112.
Although \cite{Toba_18} discussed $T_{\rm dust}$ qualitatively based on ratio of flux densities at the observed frame between 850 and 22 $\micron$, our dataset covering around the peak of FIR emission from dust enables us to do more quantitative discussion.

The dust temperature derived by the SED fitting is $T_{\rm dust}$ = 89 $\pm$ 3 K.
This is significantly higher than submillimeter galaxies (SMGs) \citep{Chapman,Kovacs} and that of FIR-selected HyLIRGs \citep{Yang} whose $T_{\rm dust}$ ranges from 30 to 60 K.
This result is consistent with what was reported qualitatively in \cite{Toba_18}.

This high dust temperature was also reported in a nearby ultraluminous IR galaxy (ULIRG), Arp 220 at $z$ = 0.018 \citep{Wilson}; the estimated $T_{\rm dust}$ of the eastern part in Arp 220 is about 80 K.
The IR luminosity surface density of WISE1013+6112 (with an effective radius of $\sim$2 kpc) is about $10^{12}$ $L_{\sun}$ kpc$^{-2}$ that is roughly consistent with that of eastern ``nucleus (0.08 kpc$\times$ 0.12 kpc)'' of Arp 220 \citep{Wilson}.
This result could suggest that WISE1013+6112 has an extreme activity that is comparable to nucleus activity of nearby ULIRGs, over the galaxy scale.

\begin{figure}
    \centering
    \includegraphics[width=0.45\textwidth]{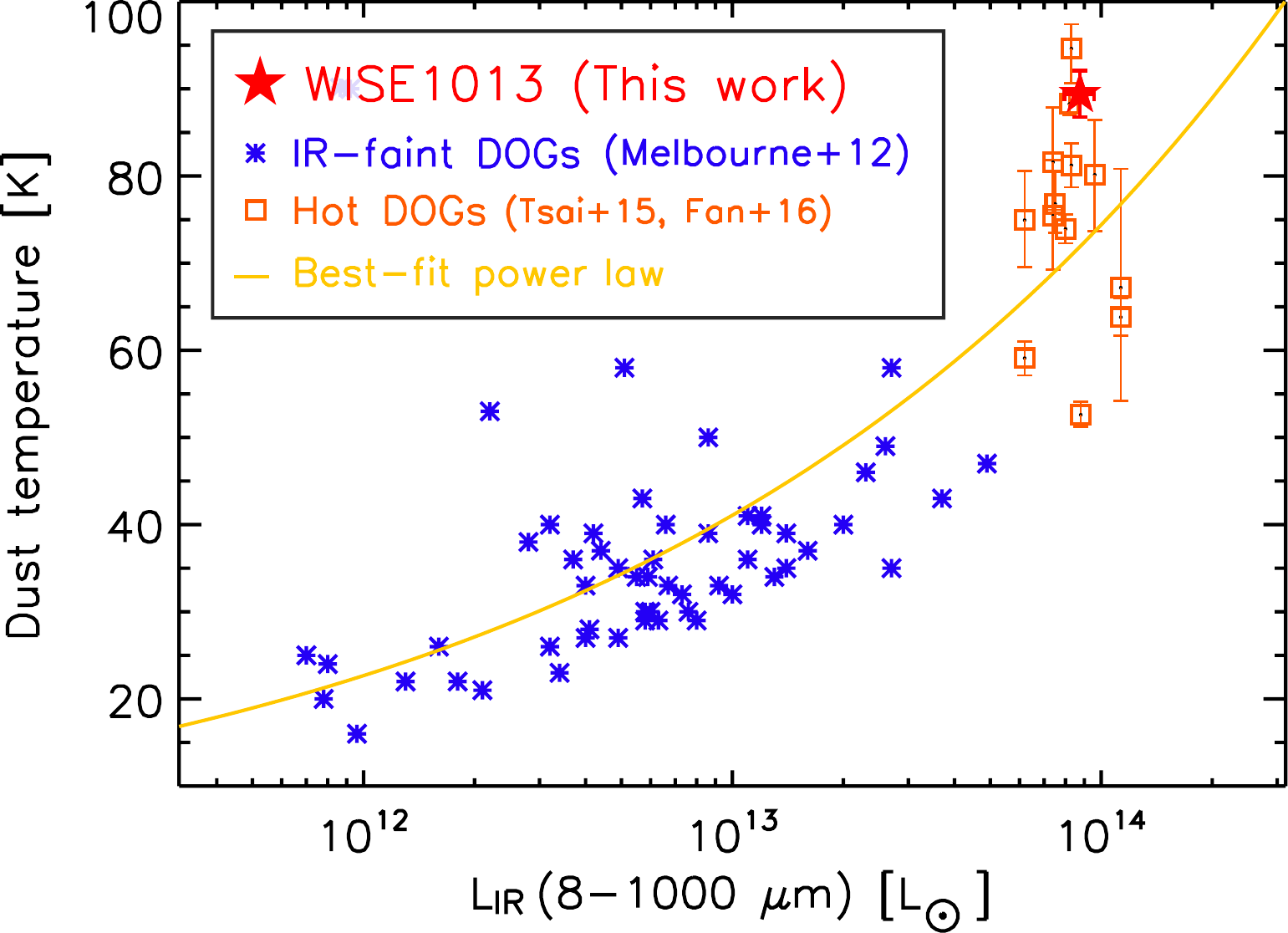}
\caption{Dust temperature as a function of IR luminosity, $L_{\rm IR}\,$(8--1000 $\micron$). Blue asterisks represent IR-faint DOGs \citep{Melbourne} while orange squares represent hot DOGs \citep{Tsai,Fan_16}. Red star represents WISE1013+6112. The yellow line is the best-fit power law for all data points, with $T_{\rm dust} = 0.02 \times L_{\rm IR}^{0.26}$.}
\label{IRdust}
\end{figure}

Figure \ref{IRdust} shows dust temperature as a function of IR luminosity for various DOG populations; IR-faint DOGs at $0.82 < z < 4.41$ \citep{Melbourne} whose MIR flux densities are fainter than those of IR-bright DOGs, hot DOGs at $1.68 < z < 4.59$ \citep{Tsai,Fan_16}, and WISE1013+6112 at $z = 3.70$.
We note that the definition of IR luminosity is often different in the literature. 
Historically, IR luminosity is defined as the one integrated over a wavelength range of 8--1000 $\micron$ \citep[e.g.,][]{Sanders,Chary}, which allows stellar emissions to contribute towards the IR luminosity.
On the other hand, recent SED fitting codes such as {\tt CIGALE} and {\tt MAGPHYS} \citep[multiwavelength analysis of galaxy physical properties;][]{da_Cunha,da_Cunha15} employ physically-motivated IR luminosity without any boundary for the integration range in wavelength. 
The IR luminosity is defined as the energy re-emitted by dust that absorbs radiations from stellar and AGNs \citep[see][]{Boquien}.
In order to compare IR luminosity with the literature under the same conditions, we integrated the best-fit SED to estimate $L_{\rm IR}$\,(8--1000 $\micron$) to be $9.0 \times 10^{13}$ $L_{\sun}$, which is plotted in Figure \ref{IRdust}.

The dust temperature of IR-faint DOGs in \cite{Melbourne} is 20--60 K that is consistent with other studies on IR-faint DOGs \citep{Calanog}.
We found that there is a correlation between $T_{\rm dust}$ and $L_{\rm IR}$\,(8--1000 $\micron$), with $T_{\rm dust} = 0.02 \times L_{\rm IR}^{0.26}$, and WISE1013+6112 is located at the luminous-end of the correlation.
The $T_{\rm dust}$--$L_{\rm IR}$ correlation for IR galaxies was reported by several authors \citep[e.g.,][]{Dunne,Chapman03,Amblard,Hwang,Magnelli,Liang}, although the origin of this correlation is still under debate \citep[see][and references therein]{Schreiber}.
One possibility is that as (i) the dust temperature is also likely to depend on redshift \citep[e.g.,][]{Magdis,Genzel,Bethermin} and (ii) those DOGs plotted in Figure \ref{IRdust} are flux--limited samples (i.e., IR luminosity correlates with redshift), the observed $T_{\rm dust}$--$L_{\rm IR}$ correlation might be due to the selection effect.

It should be noted that WISE1013+6112 with flux density at 3.4 $\micron$ $>$ 50 $\mu$Jy does not satisfy the selection criteria of hot DOGs\footnote{The exact criteria of hot DOGs are W1$>$ 17.4 ($<$ 34 $\mu$Jy), and either (i) W4$<$ 7.7 ($>$ 6.9 mJy) and W2 -- W4$>$ 8.2 or (ii) W3$<$ 10.6 ($>$ 1.7 mJy) and W2 -- W3$>$ 5.3, where W1, W2, W3, and W4 are given in Vega magnitude \citep{Eisenhardt}.} whose flux density at 3.4 $\micron$ must be smaller than 34 $\mu$Jy \citep{Eisenhardt}.
The name ``hot'' DOGs was originated from a fact that the dust temperature of hot DOGs is much hotter than that of IR-faint, classical DOGs \cite[see ][]{Dey, Wu}, which is consistent with the trend seen in Figure \ref{IRdust}.
This result could suggest that once IR luminosity exceeds $10^{14}$ $L_{\sun}$, i.e., in ELIRG regime, MIR-selected objects may have a high dust temperature regardless of satisfying the hot DOGs criteria.

\subsection{Dust mass}
\label{mass}

Finally, we derive the dust mass ($M_{\rm dust}$) of WISE1013+6112 in the same manner as in \cite{Toba_17d} where $M_{\rm dust}$ is derived from the following formula:
\begin{equation}
M_{\rm dust} = \frac{D_L^2}{1+z}\, \frac{S (\nu_{\rm obs})}{\kappa_{\rm rest} B (\nu_{\rm rest}, T_{\rm dust})},
\end{equation}
where $S (\nu_{\rm obs})$ is flux density at observed frequency ($\nu_{\rm obs}$), $D_L$ is the luminosity distance, $\kappa_{\rm rest}$ is the dust mass absorption coefficient at rest frequency ($\nu_{\rm rest}$), and $B (\nu_{\rm rest}, T_{\rm dust}$) is the Planck function at temperature $T_{\rm dust}$ and at $\nu_{\rm rest}$.
We estimated dust mass at 850 $\micron$ ($\nu_{\rm rest}$ = 353 GHz) using a dust absorption coefficient of $\kappa$ (850 $\micron$) =
0.383 cm$^2$ g$^{-1}$ \citep{Draine} and the $T_{\rm dust}$ = 89 K (see Section \ref{dust}).
Here we employed the Monte Carlo technique to calculate the dust mass and its uncertainty.
Assuming a Gaussian distribution with a mean ($T_{\rm dust}$) and sigma (its uncertainty), we randomly chose one value among the distributions as an adopted $T_{\rm dust}$.
We repeated this process 10,000 times and calculated the mean and standard deviation of the resultant $M_{\rm dust}$ distribution.
The estimated dust mass is $M_{\rm dust} = (2.1 \pm 0.1) \times 10^{8} $$M_{\sun}$.
The estimated dust-to-stellar mass ratio of WISE1013+6112 is $\log \,(M_{\rm dust}/M_{*}) = -2.96$ that is roughly consistent with that of star-forming galaxies at $z > 2.5$ \citep{Santini,Calura}.

We found that the resultant dust mass of WISE1013+6112 is inconsistent with what expected from $M_{\rm dust}$--SFR relation for local galaxies at $z < 0.3$ \citep{da_Cunha_10}; extrapolating the relation to high SFR shows that observed dust mass is about two orders of magnitude smaller than predicted \citep[see also][]{Lianou}.
This would indicate that $M_{\rm dust}$--SFR relation depend on the redshift.
This discrepancy was also reported by \cite{Hjorth} who mentioned that a difference of evolutionally sequence causes galaxies to move around in the diagram and contributes to the scatter of the $M_{\rm dust}$--SFR relation.
Indeed, give a high SFR, dust mass of dusty starburst galaxies at $z \sim$ 2--4 tends to have smaller dust mass compared to local SDSS galaxies \citep{Swinbank}.
The dust mass of those high-z starburst galaxies is roughly consistent with that of WISE1013+6112.
Nevertheless, in order to explain such a large dust mass of WISE1013+6112 at $z = 3.7$, an efficient and rapid dust formation process may be required \citep{Hjorth}.

\section{Summary and conclusions}
In this paper, we report FIR properties of an extremely-luminous DOG (WISE1013+6112) at $z_{\rm spec}$ = 3.703. 
Thanks to the multi-wavelength data set of the SDSS, {\it WISE}, SOFIA, SCUBA-2, and SMA, we pinned down their SED
at rest-frames of 0.1--300 $\micron$.
In particular, adding the observed-frames 89 and 154 $\micron$ data taken by HAWC+ is crucial to constrain the peak of FIR SED. 
We derived the physical quantities of WISE1013+6112 such as IR luminosity and dust temperature based on the SED fitting with {\tt CIGALE}.
The resultant IR luminosity is $L_{\rm IR}$ = (1.62 $\pm$ 0.08) $\times 10^{14}$$L_{\sun}$, making it one of the most luminous IR galaxies in the universe.
The derived dust temperature is $T_{\rm dust}$ = 89 $\pm$ 3 K that is significantly higher than that of other populations such as SMGs and FIR-selected galaxies.
We observed that there exists a positive correlation between $L_{\rm IR}$ and $T_{\rm dust}$ of DOGs including classical IR-faint DOGs and hot DOGs, with $T_{\rm dust} = 0.02 \times L_{\rm IR}^{0.26}$, and WISE1013+6112 is located at the luminous-end of this correlation.
The dust mass inferred from $T_{\rm dust}$ is $M_{\rm dust} = (2.1 \pm 0.1) \times 10^{8}$ $M_{\sun}$ that is  inconsistent with what expected from $M_{\rm dust}$--SFR relation for local galaxies. 
An efficient formation of dust from the metals may need to be considered to produce such a high dust mass given the redshift of $z = 3.7$.

\acknowledgments
We gratefully acknowledge the anonymous referee for a careful reading of the manuscript and very helpful comments.
We are deeply thankful to Drs. Lopez-Rodrigue Enrique and Ralph Y. Shuping for helping with the observations and data analysis taken by HAWC+/SOFIA.
We also thank Prof. Denis Burgarella for helping us to understand {\tt CIGALE} code.

Based [in part] on observations made with the NASA/DLR Stratospheric Observatory for Infrared Astronomy (SOFIA). SOFIA is jointly operated by the Universities Space Research Association, Inc. (USRA), under NASA contract NNA17BF53C, and the Deutsches SOFIA Institut (DSI) under DLR contract 50 OK 0901 to the University of Stuttgart.

The Submillimeter Array is a joint project between the Smithsonian Astrophysical Observatory and the Academia Sinica Institute of Astronomy and Astrophysics and is funded by the Smithsonian Institution and the Academia Sinica.

The James Clerk Maxwell Telescope has historically been operated by the Joint Astronomy Centre on behalf of the Science and Technology Facilities Council of the United Kingdom, the National Research Council of Canada and the Netherlands Organisation for Scientific Research.
Additional funds for the construction of SCUBA-2 were provided by the Canada Foundation for Innovation. 
Funding for SDSS-III has been provided by the Alfred P. Sloan Foundation, the Participating Institutions, the National Science Foundation, and the U.S. Department of Energy Office of Science. The SDSS-III web site is http://www.sdss3.org/.

Funding for SDSS-III has been provided by the Alfred P. Sloan Foundation, the Participating Institutions, the National Science Foundation, and the U.S. Department of Energy Office of Science. The SDSS-III web site is http://www.sdss3.org/.
SDSS-III is managed by the Astrophysical Research Consortium for the Participating Institutions of the SDSS-III Collaboration including the University of Arizona, the Brazilian Participation Group, Brookhaven National Laboratory, Carnegie Mellon University, University of Florida, the French Participation Group, the German Participation Group, Harvard University, the Instituto de Astrofisica de Canarias, the Michigan State/Notre Dame/JINA Participation Group, Johns Hopkins University, Lawrence Berkeley National Laboratory, Max Planck Institute for Astrophysics, Max Planck Institute for Extraterrestrial Physics, New Mexico State University, New York University, Ohio State University, Pennsylvania State University, University of Portsmouth, Princeton University, the Spanish Participation Group, University of Tokyo, University of Utah, Vanderbilt University, University of Virginia, University of Washington, and Yale University.

This publication makes use of data products from the {\it Wide-field Infrared Survey Explorer}, which is a joint project of the University of California, Los Angeles, and the Jet Propulsion Laboratory/California Institute of Technology, funded by the National Aeronautics and Space Administration.

This work is supported by JSPS KAKENHI Grant numbers 18J01050 and 19K14759 (Y.Toba), 16H03958, 17H01114, and 19H00697 (T.Nagao), and 17K05384 (Y.Ueda).
Y.Toba and W.H.Wang acknowledge the support from the Ministry of Science and Technology of Taiwan (MOST 105-2112-M-001-029-MY3). 

\vspace{5mm}
\facilities{SOFIA (HAWC+), Sloan, {\it WISE}, SMA, JCMT}


\software{IDL, IDL Astronomy User's Library \citep{Landsman}, CASA (v5.5.0) \citep{McMullin}, {\sf CIGALE} \citep{Boquien}}


\end{document}